\documentstyle[psfig,multicol,aps]{revtex}
\begin{document}
\draft
\title{Inverse Amplitude Method and Chiral Zeros}
\author{Torben Hannah\cite{THannah}}
\address{Nordita, Blegdamsvej 17, DK-2100 Copenhagen {\O}, Denmark}
\maketitle
\begin{abstract}
The inverse amplitude method has previously been successfully
applied to $\pi\pi$ scattering in order to extend the range
of applicability of chiral perturbation theory. However, in
order to take the chiral zeros into account systematically,
the previous derivation of the inverse amplitude method has
to be modified. It is shown how this can be done to both one
and two loops in the chiral expansion. In the physical region,
the inclusion of these chiral zeros has very little significance,
whereas they become essential in the sub-threshold region.
Finally, the crossing properties of the inverse amplitude method
are investigated in the sub-threshold region.
\end{abstract}

\begin{multicols}{2}

Chiral perturbation theory (ChPT) \cite{ref:We79,ref:GL84}
has become a very successful methodology for low-energy hadron
physics. Within this methodology one obtains a systematic
expansion in powers of external momenta and light quark masses, or
equivalently in the number of loops. However, unitarity is only
satisfied perturbatively in the chiral expansion, which gives a
severe restriction on the applicability of ChPT. Therefore, in order
to extend the validity of ChPT, several methods have been proposed to
combine exact unitarity and the chiral expansion.
One such method is the inverse amplitude method (IAM), which has
been analyzed to both one-loop \cite{ref:Tru88} and two-loop
\cite{ref:Han97} order in the chiral expansion. This has shown
that the IAM is indeed a successful and systematic method to extend
the range of applicability of ChPT.

However, the IAM has to be somewhat modified when there are zeros in
the amplitudes \cite{ref:BP96}. This is the case for $\pi\pi$ scattering
where chiral dynamics demand that the $S$ waves have zeros below threshold,
whereas $P$ and higher partial waves have fixed kinematical zeros at
threshold. The previous derivation of the IAM \cite{ref:Tru88,ref:Han97}
has been based upon the assumption that these zeros occur at the same
energy as for the lowest order ChPT result. This assumption is in fact true
for the $P$ and higher partial waves, whereas the same is not necessarily
the case for the $S$ waves. Thus, the derivation of the IAM should be
somewhat modified in the case of the $S$ waves in order to systematically
account for the occurrence of the chiral zeros. In this brief report
it is shown how this is possible to both one and two-loop order in
the chiral expansion. Since the main interest will be in the sub-threshold
behavior, the following derivation of this generalized IAM will be
restricted to the elastic approximation.

The elastic unitarity relation for the $\pi\pi$ partial waves $t$
is given by
\begin{equation}
\label{eq:uni}
{\rm Im}t^I_l(s)=\sigma (s)|t^I_l(s)|^2 ,
\end{equation}
where $s$ is the square of the c.m. energy, $I$ and $l$ are
isospin and angular momentum indices, and $\sigma$ is the
phase-space factor. In ChPT the elastic $\pi\pi$ partial waves are
now known to two loops in the chiral expansion \cite{ref:KMSF95,ref:Bij96}.
These partial waves are given by the following expansion
\begin{equation}
\label{eq:expar}
t^I_l(s) = t^{I(0)}_l(s)+t^{I(1)}_l(s)+t^{I(2)}_l(s) ,
\end{equation}
where $t^{(0)}$ is the lowest order result, $t^{(1)}$ is the
one-loop correction, and $t^{(2)}$ the additional two-loop
correction. They satisfy the elastic unitarity relation (\ref{eq:uni})
perturbatively
\begin{eqnarray}
\label{eq:puni}
{\rm Im}t^{I(0)}_l(s) & = & 0 , \nonumber \\
{\rm Im}t^{I(1)}_l(s) & = & \sigma (s){t^{I(0)}_l}^2(s) ,
\nonumber \\
{\rm Im}t^{I(2)}_l(s) & = & \sigma (s)2t^{I(0)}_l(s){\rm Re}
t^{I(1)}_l(s) .
\end{eqnarray}
Since perturbative unitarity only works well rather close to
threshold, this will give a severe restriction on the
applicability of ChPT. However, with the use of the IAM,
the range of applicability of ChPT can be substantially extended.
The previous starting point for this method \cite{ref:Tru88,ref:Han97}
was to write down a dispersion relation for the function
$\Gamma ={t^{(0)}}^2/t$. However, in order to systematically
account for the occurrence of the chiral zeros, one has to modify
the function $\Gamma$ and now write it as $\Gamma =(t^{(0)}+\epsilon )^2/t$.
The $\epsilon$ parameter parameterizes the change in the position of the
zero in the partial wave $t$ away from the lowest order result, i.e.,
$\epsilon$ is determined by the equation
\begin{equation}
\label{eq:ep}
t^{I(0)}_l(s_z)+\epsilon^I_l = 0 ,
\end{equation}
where $s_z$ is the position of the zero in the partial wave $t$. Of
course, the precise value of $s_z$ and therefore also the value of
$\epsilon$ is unknown, but they may both be approximated by the use
of the chiral expansion, which should apply unambiguously in the
sub-threshold region. Unfortunately, the zero in the chiral
expansion can only be found analytically to lowest order. However,
the zero may be determined very accurately to a given order in the
chiral expansion by solving the equation
$t^{(0)}(s)+t^{(1)}(s^{(0)}_z)+t^{(2)}(s^{(0)}_z)+\cdots =0$,
where $s^{(0)}_z$ is the position of the zero in the lowest order
result $t^{(0)}$. This has been checked to work very well indeed
for both one and two-loop ChPT. Therefore, comparing with Eq.
(\ref{eq:ep}) one finds that $\epsilon$ may be expanded in ChPT as
\begin{equation}
\label{eq:epex}
\epsilon^I_l = \epsilon^{I(1)}_l+\epsilon^{I(2)}_l+\cdots ,
\end{equation}
where
\begin{equation}
\label{eq:epdef}
\epsilon^{I(1)}_l = t^{I(1)}_l(s^{(0)}_z) ,\;\;
\epsilon^{I(2)}_l = t^{I(2)}_l(s^{(0)}_z) ,\;\;\cdots .
\end{equation}
Having expanded $\epsilon$ in ChPT it is now possible to obtain the
function $\Gamma$ to a given order in the chiral expansion. To two
loops one finds that the result $\Gamma^{(2)}$ is given by
\begin{eqnarray}
\label{eq:exgamma}
\Gamma^{I(2)}_l(s) = t^{I(0)}_l(s)-t^{I(1)}_l(s)+
\frac{{t^{I(1)}_l}^2(s)}{t^{I(0)}_l(s)}-t^{I(2)}_l(s) \nonumber \\
+2\epsilon^{I(1)}_l\left( 1-\frac{t^{I(1)}_l(s)}{t^{I(0)}_l(s)}
\right)
+2\epsilon^{I(2)}_l+\frac{{\epsilon^{I(1)}_l}^2}{t^{I(0)}_l(s)} .
\end{eqnarray}
This expression for $\Gamma^{(2)}$ is well-defined at $s^{(0)}_z$
where it coincides with the truncation of the chiral expansion
(\ref{eq:expar}). In fact, it is analytic in the whole complex
$s$ plane with cuts for $-\infty <s<0$ and $4M_{\pi}^2<s<\infty$.
Therefore, it is possible to write down a dispersion relation for
$\Gamma^{(2)}$ with four subtractions in order to ensure the
convergences of the integrals. For the right cut perturbative
unitarity (\ref{eq:puni}) gives
${\rm Im}\Gamma^{(2)}=-\sigma ({t^{(0)}}^2+2\epsilon^{(1)}t^{(0)})$,
whereas the left cut is simply given by ${\rm Im}\Gamma^{(2)}$.

Turning to the original function $\Gamma =(t^{(0)}+\epsilon )^2/t$,
it is known from the fundamental principles of S matrix theory that
the partial wave $t$ is analytic in the complex $s$ plane with a
right cut for $4M_{\pi}^2<s<\infty$ and a left cut for
$-\infty <s<0$. Assuming that the only zero in $t$ is the one
below threshold demanded by chiral dynamics, the function $\Gamma$
is also analytic in the whole complex $s$ plane with the same cut
structure as $t$. It is therefore possible to write down a
dispersion relation for $\Gamma$ in the same way as
for $\Gamma^{(2)}$. In this case, however, exact unitarity is used
in order to compute the right cut, i.e., Eq. (\ref{eq:uni}) gives
${\rm Im}\Gamma = -\sigma (t^{(0)}+\epsilon )^2$.
Since the precise value of $\epsilon$ is unknown, the terms
involving $\epsilon$ have to be expanded in ChPT using Eq.
(\ref{eq:epex}). This gives
${\rm Im}\Gamma = -\sigma ({t^{(0)}}^2+2\epsilon^{(1)}t^{(0)})$
to two loops in the chiral expansion, which is used for the
right cut in the dispersion relation for $\Gamma$.

The left cut and the subtraction constants cannot be computed in the same way,
but it is possible to evaluate them to a given order in the chiral
expansion. Evaluating both the left cut and the subtraction constants
to two loops in the chiral expansion, one finds that the dispersion
relation for $\Gamma$ is exactly the same as the dispersion relation
for $\Gamma^{(2)}$. Therefore, one has the following relation
\begin{equation}
\label{eq:dispIAM}
\frac{\left( t^{I(0)}_l(s)+\epsilon^I_l\right)^2}{t^I_l(s)} =
\Gamma^{I(2)}_l(s) , 
\end{equation}
from where it is possible to determine the position of
the zero in $t$ by setting the right hand side equal to zero
and solving for $s$. One can then determine $\epsilon$
from Eq. (\ref{eq:ep}) with a result that should of course be
close to the value obtained from the chiral expansion to two
loops (\ref{eq:epex}). From Eq. (\ref{eq:dispIAM}) one finds
that the final result can be written in the form
\begin{equation}
\label{eq:IAM2}
t^I_l(s) = \frac{\left( t^{I(0)}_l(s)+\epsilon^I_l\right)^2}
{\Gamma^{I(2)}_l(s)} .
\end{equation}
An important property of this generalized IAM 
is that it coincides with the chiral expansion (\ref{eq:expar})
up to two-loop order. This is due to the fact that the left cut,
the subtraction constants, and $\epsilon$ in the dispersion
relation for the IAM have been evaluated to two-loop order in
ChPT. If the left cut had been approximated by something else
\cite{ref:BP96}, the IAM would not have been consistent with
the chiral expansion (\ref{eq:expar}) up to two-loop order.
Furthermore, the generalized IAM will satisfy unitarity exactly
up to two loops in $(t^{(0)}+\epsilon )^2$, which should work
extremely well due to the suppression of the higher order contributions.
Finally, with the assumption that $\epsilon$ to all orders in ChPT is
exactly zero, one finds that Eq. (\ref{eq:IAM2}) coincides with the [0,2]
Pad\'{e} approximant previously derived in Ref. \cite{ref:Han97}. Since
this assumption is in fact true for the $P$ partial wave, one finds that
the generalized IAM in this case coincides with the previously obtained
result.

The generalized IAM can in general be applied to any given order
in the chiral expansion and therefore also to the one-loop
approximation. In this case one finds the result
\begin{equation}
\label{eq:IAM1}
t^I_l(s) = \frac{\left( t^{I(0)}_l(s)+\epsilon^I_l\right)^2}
{\Gamma^{I(1)}_l(s)} ,
\end{equation}
where $\Gamma^{(1)}$ is given by
\begin{equation}
\label{eq:exgamma1}
\Gamma^{I(1)}_l(s) = t^{I(0)}_l(s)-t^{I(1)}_l(s)+
2\epsilon^{I(1)}_l .
\end{equation}
Expanding this result to one loop in the chiral expansion, one
finds that it coincides with one-loop ChPT. Furthermore,
neglecting the $\epsilon$ terms one finds the [0,1] Pad\'{e}
approximant, which has previously been extensively analyzed
\cite{ref:Tru88}.

In the following the pion mass $M_{\pi}$ and the pion decay constant
$F_{\pi}$ are set equal to $M_{\pi}=139.6$ MeV and
$F_{\pi}=92.4$ MeV, respectively. In addition, the chiral expansion
is given in terms of a number of low-energy constants, which have
to be determined before it is possible to obtain numerical results
for the IAM. The low-energy constants occurring in the IAM to one
and two loops have previously been determined without taking the
chiral zeros into account \cite{ref:Han97}. Thus, in order to
show how important the inclusion of the chiral zeros is, the same
values of the low-energy constants are also used in the generalized
IAM to one and two loops.

\begin{figure}
\centerline{\psfig{figure=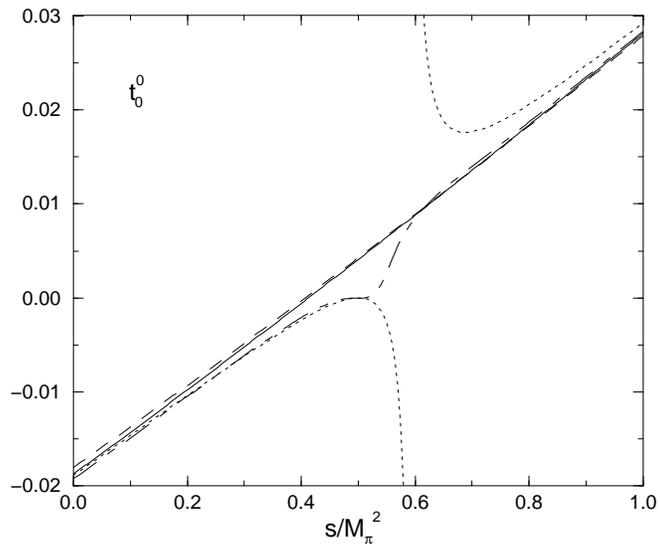,width=9cm,angle=-90}}
\caption{The partial wave $t^0_0$ in the region
$0\leq s\leq M^2_{\pi}$. The solid line is the generalized IAM
to two loops, the dashed line the generalized IAM to one loop,
the dashed-dotted line the IAM to two loops without taking the
chiral zeros into account, and the dotted line the IAM to one loop
without taking the chiral zeros into account.}
\label{Fig1}
\end{figure}

\begin{figure}
\centerline{\psfig{figure=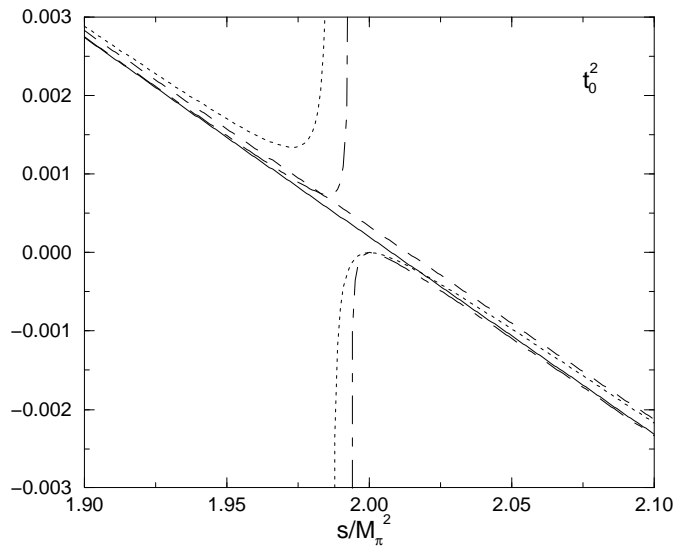,width=9cm,angle=-90}}
\caption{The partial wave $t^2_0$ in the region
$1.9M^2_{\pi}\leq s\leq 2.1M^2_{\pi}$. The curves are as in
Fig. \protect\ref{Fig1}.}
\label{Fig2}
\end{figure}

With the low-energy constants fixed at the
values given in Ref. \cite{ref:Han97}, the parameter $\epsilon$ in the
generalized IAM to one and two loops can be determined with
results that are indeed close to the chiral expansion of
$\epsilon$ to one and two loops, respectively. Thus, with
all the parameters in the generalized IAM fixed, one can compare
with the results obtained previously without taking the occurrence
of the chiral zeros into account \cite{ref:Han97}. In the physical
region one finds that the two approaches are practical identical,
indicating that the inclusion of the chiral zeros in the IAM
has very little significance in the physical region. However,
in the sub-threshold region, the inclusion of these zeros becomes
essential as shown in Figs. \ref{Fig1} and \ref{Fig2}. Indeed,
the generalized IAM generates the chiral zeros, whereas the IAM
without taking the chiral zeros into account clearly fails in the
displayed energy regions.

An important constraint on the $\pi\pi$ scattering amplitude is
due to crossing symmetry. Of course ChPT satisfies this symmetry
exactly, whereas the same is not necessarily true in the case of
the IAM. Therefore, it would be interesting to investigate how well
the IAM actually satisfies crossing. In order to do this one needs
to express the consequences of crossing in terms of a finite number
of partial waves. This is indeed possible in the sub-threshold
region where one has so-called crossing sum rules \cite{ref:Ros70}.
These sum rules provide a necessary and sufficient set of conditions
for crossing. There are five sum rules involving only $S$ and $P$
partial waves, which might be written generically as
\begin{equation}
\label{eq:cros}
\int_0^{4M_{\pi}^2}ds\;\omega_l(s)\sum_I\alpha_It^I_l(s) = 0 ,
\end{equation}
where the $\alpha$ are constants. The explicit forms of these
crossing sum rules can be obtained from the appendix in Ref.
\cite{ref:BP96}. In order to evaluate how well crossing is
satisfied, the following measure is used
\begin{equation}
\label{eq:tcros}
100\times\frac{|\int_0^{4M_{\pi}^2}ds\;\omega_l(s)\sum_I
\alpha_It^I_l(s)|}{\int_0^{4M_{\pi}^2}ds\;|\omega_l(s)\sum_I
\alpha_It^I_l(s)|} .
\end{equation}
Since the crossing sum rules involve the partial waves in the
sub-threshold region, where the inclusion of the chiral zeros in
the IAM becomes essential, the measure (\ref{eq:tcros}) is only
evaluated for the generalized IAM with the results given in Table
\ref{Table1}. From this table it is observed that the IAM to two
loops significantly improves the crossing symmetry of the one-loop
approximation. In fact, crossing is satisfied very well indeed in
the two-loop approach, the violation being less than 0.2\% in all
cases. For the one-loop approximation the violation is larger but
still less than 1.3\% in all cases.

The crossing symmetry of the IAM can also be investigated with the
use of Martin inequalities \cite{ref:JM64,ref:Auber70}. These
inequalities are exact consequences of the general principles of
quantum field theory, which mainly consist in the constraints coming
from analyticity, crossing symmetry, and positivity of absorptive
parts. A very large number of inequalities has been derived, but the
most interesting are for the $\pi^0\pi^0$ $S$ partial wave, which is
given by
\begin{equation}
\label{eq:t000}
t^{00}_0(s) = \mbox{$\frac{1}{3}$}\left( t^0_0(s)+2t^2_0(s)
\right) .
\end{equation}
With $s$ given in units of $M^2_{\pi}$, one has the following
inequalities on the derivatives of $t^{00}_0$
\cite{ref:JM64}
\begin{eqnarray}
\label{eq:dt0in}
\frac{dt^{00}_0(s)}{ds} & < & 0\;\; {\rm for}\;\; 0<s<1.219 ,
\nonumber \\
\frac{dt^{00}_0(s)}{ds} & > & 0\;\; {\rm for}\;\; 1.697<s<4 ,
\nonumber \\
\frac{d^2t^{00}_0(s)}{ds^2} & > & 0\;\; {\rm for}\;\; 0<s<1.7,
\end{eqnarray}
which implies that $t^{00}_0$ has a minimum in the interval
(1.219,1.697). In addition, the following inequalities have
also been obtained \cite{ref:JM64}
\begin{eqnarray}
\label{eq:t0in}
& t^{00}_0 & (4) > t^{00}_0(0) > t^{00}_0(2[1+1/\sqrt{3}]) ,
\nonumber \\
& t^{00}_0 & (3.205) > t^{00}_0(0.2134) > t^{00}_0(2.9863) .
\end{eqnarray}
One finds that the axiomatic
constraints (\ref{eq:dt0in}) and (\ref{eq:t0in}) are satisfied for
the generalized IAM to both one and two loops. In the two-loop approach
the minimum occurs at $s/M^2_{\pi}=1.652$, whereas in the one-loop
approximation one obtains the minimum at $s/M^2_{\pi}=1.643$.
If the neutral pion mass $M_{\pi}=135.0$ MeV had been used
instead of the charge pion mass $M_{\pi}=139.6$ MeV, the
result would have been slightly different. However, the axiomatic
constraints are also satisfied in this case.

Inequalities involving the definite isospin $S$ and $P$ partial waves
have also been obtained. For instance, one has the following axiomatic
constraints \cite{ref:Auber70}
\begin{eqnarray}
\label{eq:tiin}
&& at^0_0(0)+bt^2_0(0)+ct^1_1(0) < At^0_0(4)+Bt^2_0(4) ,
\nonumber \\
&& t^0_0(s)-t^2_0(s)+9\cos\theta_st^1_1(s) > \nonumber \\
&& \;\;\;\;\;\;\;\;\;\;\;\;\;\;\;\;\;\;\;\;\;\;\;\;\;\;
t^0_0(t)-t^2_0(t)-9\cos\theta_tt^1_1(t) , \nonumber \\
&& t^0_0(s)-t^2_0(s)+3\cos\theta_st^1_1(s)
\mbox{ \raisebox{-0.8ex}{$\stackrel{\displaystyle >}{<}$} } \nonumber \\
&& \;\;\;\;\;\;\;\;\;\;\;\;\;\;\;\;\;\;\;\;\;
\mbox{$\frac{1}{3}$}t^0_0(t)+\mbox{$\frac{2}{3}$}
t^2_0(t)-6\cos\theta_tt^1_1(t) ,
\end{eqnarray}
where the specified values of $s$ and $t$ together with the
values of the coefficients are given in Ref. \cite{ref:Auber70}.
The inequalities (\ref{eq:tiin}) give altogether 27 different
constraints on the partial waves, which have been tested in the
case of the generalized IAM. For the two-loop approach, one finds
that all the above constraints are satisfied, whereas the one-loop
approximation slightly violates two of the constraints. This
agrees nicely with the previous analysis based on the crossing sum
rules, where it was shown that the generalized IAM to two loops
improves the crossing symmetry of the one-loop approximation. This
conclusion is not likely to change if even other axiomatic
constraints had been tested.

To summarize, it has been shown how the IAM can be generalized in order
to systematically take the chiral zeros into account. In the physical
region one finds that the inclusion of these zeros has very
little significance, which justify the previous neglect of these
zeros. However, below threshold it becomes essential to include
the chiral zeros in the derivation of the IAM. Furthermore, the
crossing symmetry of the IAM is investigated by the use of
sub-threshold crossing sum rules and Martin inequalities. It is found
that the generalized IAM to two loops satisfies crossing symmetry
very well indeed, whereas the violation of crossing is somewhat
larger for the one-loop approximation.

Recently, the IAM has been applied to coupled channels \cite{ref:Oller98}
in order to extend the applicability of the single channel IAM
to even higher energies. However, this coupled channel IAM generates
poles in the sub-threshold region just as the previous derived single
channel IAM did. Thus, the derivation of the coupled channel IAM should
in principle be somewhat modified in order to remove these poles. However,
since the poles in the single channel IAM can be removed without any
significant influence in the physical region, the same in also likely
to be true in the coupled channel case.

\end{multicols}

\newpage

\begin{table}
\caption{The measure Eq. (\protect\ref{eq:tcros}) for the
generalized IAM to one loop (GIAM1) and for the generalized
IAM to two loops (GIAM2). The numbers refer to the corresponding
crossing sum rule as given in Ref. \protect\cite{ref:BP96}.}
\label{Table1}
\begin{tabular}{cccccc}
&1&2&3&4&5\\
\tableline
GIAM1&0.39&1.25&0.57&0.18&0.55\\
GIAM2&0.06&0.18&0.07&0.04&0.09\\
\end{tabular}
\end{table}

\end{document}